# Pluto's interaction with its space environment: Solar Wind, Energetic Particles & Dust

F. Bagenal[1], M. Horányi[1], D. J. McComas[2,3], R. L. McNutt, Jr.[4], H. A. Elliott[2], M. E. Hill[4], L. E. Brown[4], P. A. Delamere[5], P. Kollmann[4], S. M. Krimigis[4,6], M. Kusterer[4], C. M. Lisse[4], D. G. Mitchell[4], M. Piquette[1], A. R. Poppe[7], D. F. Strobel[8], J. R. Szalay[1], P. Valek[2], J. Vandegriff[4], S. Weidner[2], E. J. Zirnstein[2], S. A. Stern[9], K. Ennico[10], C. B. Olkin[9] H. A. Weaver[4], L. A. Young[9]

[1]Laboratory of Atmospheric and Space Physics,
University of Colorado, Boulder, CO 80600, USA
[2]Southwest Research Institute, San Antonio, TX 78228 USA
[3]University of Texas at San Antonio, San Antonio, TX 78249 USA,
[4] Johns Hopkins University Applied Physics Laboratory, Laurel, MD 20723 USA
[5]University of Alaska, Fairbanks, AK 99775 USA
[6]Academy of Athens, 28 Panapistimiou, 10679 Athens, Greece
[7]Space Sciences Laboratory, University of California, Berkeley, CA 94720, USA
[8]Johns Hopkins University, Baltimore, MD 21218 USA
[9]Southwest Research Institute, Boulder, CO 80302 USA
[10]NASA Ames Research Center, Moffett Field, CA 94035 USA

**Abstract**

**The New Horizons spacecraft carried three instruments that measured the space environment near Pluto as it flew by on 14 July 2015. The Solar Wind Around Pluto instrument revealed an interaction region confined sunward of Pluto to within ~6 Pluto radii. The surprisingly small size is consistent with a reduced atmospheric escape rate as well as a particularly high solar wind flux. The Pluto Energetic Particle Spectrometer Science Investigation (PEPSSI) observations suggested ions are accelerated and/or deflected around Pluto. In the wake of the interaction region PEPSSI observed suprathermal particle fluxes about 1/10 the flux in the interplanetary medium, increasing with distance downstream. The Student Dust Counter, which measures >1.4$\mu$m grains, detected 1 candidate impact in ±5days around closest approach, indicating an upper limit for the dust density in the Pluto system of 4.6km$^{-3}$.**

NASA's New Horizons spacecraft flew past Pluto on 14 July 2015 after a journey of over nine years [1]. Scientific objectives of the New Horizons mission include quantifying the rate at which atmospheric gases are escaping Pluto [2] and describing its interaction with the surrounding space environment. The two New Horizons instruments that measure charged particles are the Solar Wind Around Pluto (SWAP) instrument [3] and the Pluto Energetic Particle Spectrometer Science Investigation (PEPSSI) instrument [4]. The Venetia Burney Student Dust Counter (SDC) counts the micron-sized dust grains that hit the detectors mounted on the ram direction of the spacecraft [5]. This paper describes preliminary results from these three instruments from the Pluto encounter period (the geometry of which is illustrated in **Figure 1**). New Horizons reached its closest approach distance of 11.54 $R_P$, where a Pluto radius is $R_P$ = 1187 km [1], on Day Of Year (DOY) 195 at 11:49 Coordinated Universal Time (UTC).

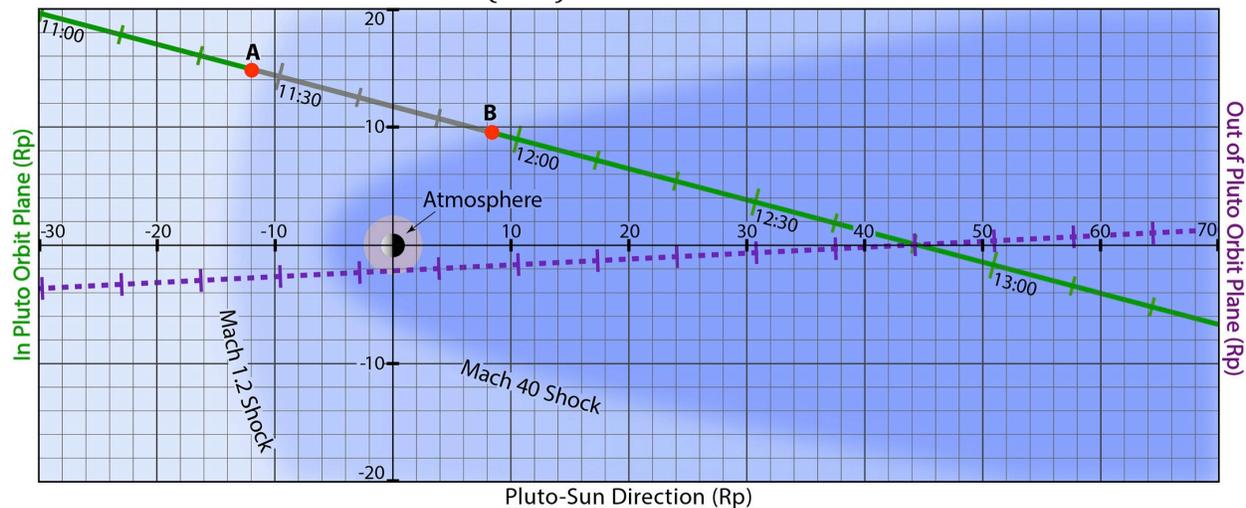

***Fig. 1: Geometry of New Horizons trajectory through the solar wind interaction with Pluto's atmosphere on DOY 195.*** *Along the grey section of the trajectory between points A and B the New Horizons spacecraft pointed the SWAP instrument FOV away from the solar direction. The pink circle shows the extent of Pluto's atmosphere [2].*

Initial studies of the solar wind interaction with Pluto's atmosphere [6, 7, 8, 9, 10, 11, 12, 13, 14, 15], all assuming the absence of an intrinsic magnetic field, suggested that it would depend on whether the atmospheric escape flux is strong – producing a "comet-like" interaction where the interaction region is dominated by ion pick-up and many times larger than the object – or weak  - producing a "Mars-like" interaction dominated by ionospheric currents with limited upstream pick-up and where the scale size is comparable to the object.  Before the New Horizons flyby the estimates of the atmospheric escape ranged from as low as 1.5 x $10^{25}$ s$^{-1}$ to as high as 2 x $10^{28}$ s$^{-1}$ [16, 17, 18, 19, 20, 21, 22, 23]. Combining these atmospheric escape rates with Voyager and New Horizons observations of the solar wind at 33 AU produced estimates of the scale of the interaction region from 7 to 1000 $R_P$ [24].

New Horizons flew past Pluto at a distance of 32.9 Astronomical Units (AU) from the Sun. At the time of encounter, Pluto was 1.9° above the ecliptic plane on its eccentric orbit. The flyby occurred as the Sun was in the descending phase of the solar cycle. In Table 1 we compare the interplanetary plasma conditions predicted based on Voyager data for 33 AU [24] with the observations made by the New Horizons SWAP instrument. In the absence of a direct measurement of the local magnetic field, we assume an interplanetary magnetic field (IMF) of 0.3 nT, at the upper end of the range observed by Voyager (as discussed further below).

|  |  | Predicted 10th percentile | Predicted Median | Predicted 90th Percentile | Observed by SWAP |
|---|---|---|---|---|---|
| Radial flow speed | $V_R$ [km/s] | 380 | 430 | 480 | 403 |
| Proton density | $n$ [cm$^{-3}$] | 0.0020 | 0.0058 | 0.014 | 0.025 |
| Proton flux | $nV$ [km s$^{-1}$ cm$^{-3}$] | 0.84 | 2.4 | 7.0 | 10 |
| Proton temperature | $T$ [K] <br> $T$ [eV] | 3040 <br> 0.26 | 6650 <br> 0.57 | 16800 <br> 1.5 | 7700 <br> 0.66 |
| Proton thermal pressure (IPUI pressure*) | $P = nkT$ [fPa] | 0.12 | 0.53 <br> (20 ± 8*) | 2.1 | 2.5 |
| Proton ram pressure | $P = \rho V^2$ [pPa] | 0.55 | 1.7 | 4.0 | 6.0 |
| Sound speed (with IPUIs*) | $V_s = (\gamma kT/m_i)^{1/2}$ [km/s] | 6.3 | 9.4 <br> (58*) | 15 | 10 <br> (28*) |
| Sonic Mach number (with IPUIs*) | $M_S = V_r/V_s$ | 60 | 46 <br> (7.5*) | 32 | 40 <br> (14*) |
| Magnetic field strength | $B$ [nT] | 0.08 | 0.15 | 0.28 | 0.3† |
| Pickup CH$_4^+$ gyroradius | $R_{gyro}$ [$R_P$] | 670 | 400 | 240 | 190† |
| Alfven speed | $V_A = B/(\mu_0\rho)^{1/2}$ [km/s] | 22 | 45 | 96 | 41† |
| Alfven Mach number | $M_{Alf} = V_R/V_A$ | 4.6 | 9.5 | 20 | 9.8† |
| Magnetosonic Mach No. (with IPUIs*) | $M_{MS} = V_r/(V_A^2+V_s^2)^{1/2}$ | 17 | 9 <br> (6*) | 5 | 9.5† <br> (8*†) |

**Table 1: Solar Wind Conditions at 33 AU:** predictions based on Voyager 2 plasma data obtained 1988-1992 between 25 and 39 AU [24] and observed by the New Horizons SWAP instrument.
\* Including the thermal pressure of interstellar pick-up ions (IPUIs) from [30, 31, 32].
† For the Pluto flyby we take an interplanetary magnetic field strength of ~0.3 nT.

**Solar Wind Around Pluto**

*The SWAP Instrument*

The SWAP instrument was specifically designed and optimized for the New Horizons mission, with the primary design drivers being (i) a large aperture and geometric factor to measure accurately the tenuous solar wind at ~33 AU, (ii) a minimum use of spacecraft resources such as mass and power, and (iii) mounting on a spacecraft that would rotate over a very wide range of spacecraft pointing directions throughout the flyby. Early in the mission development phase, the flyby pointing directions were not well known but the best information available at that time was that most spacecraft pointing would involve rotation around a single spacecraft axis – the spacecraft Z-axis. Thus, SWAP is designed to have an extremely broad acceptance angle (~276°) in the plane perpendicular to this axis, which requires an electrostatic top-hat style analyzer with its axis of symmetry aligned with the spacecraft's Z-axis. Ions are bent through the electrostatic analyzer, pass through a nearly field-free conical region, and are focused into a coincidence detector section, which provides very high signal-to-noise ratio measurements for solar wind ions. Details of the SWAP instrument design are provided by [3] and SWAP data has already been used to examine the Jovian magnetosphere and distant magnetotail [25, 26, 27, 28, 29].

*The Solar Wind at the Time of the Pluto Flyby*

At the time of the flyby, the solar wind conditions near Pluto measured by SWAP were nearly constant, which is advantageous for interpreting the solar wind – Pluto interaction. The top panel of **Figure 2** shows a color spectrogram of SWAP coincidence counts as a function of energy-per-charge (E/Q) and time. On the left and right sides of the plot, away from the Pluto interaction, the red and yellow bands at a little under 1 and 2 keV/q are solar wind protons and alpha particles, respectively. We calculate the solar wind proton parameters for these intervals [34] and find a strong consistency between the values before ~11:20 on DOY 195 and again on DOY 196 (black points). These intervals represent the unperturbed solar wind ahead of and beyond Pluto along the New Horizons trajectory, respectively. Interpolating between these points (black line), we infer that at the time of the New Horizons closest approach (11:48 – green vertical dashed line), a solar wind speed of ~403 km s$^{-1}$, a proton density of ~0.025 cm$^{-3}$, a proton temperature of ~7700 K (0.7 eV), a proton dynamic pressure of ~6.0 pPa, and a core solar wind proton thermal pressure of ~2.5x10$^{-3}$ pPa (Table 1). From the properties of just the thermal solar wind we calculate a sonic Mach number of ~40. The sonic Mach number is substantially reduced to 14 if we include the interstellar pick-up ions (IPUIs), which provide roughly an order of magnitude greater thermal pressure at these heliocentric distances [31, 32]. The measured unusually high solar wind density and associated pressures for this distance are likely due to a relatively strong traveling interplanetary shock that passed over the spacecraft five days earlier on DOY 190.

New Horizons was not equipped with a magnetometer, so the interplanetary magnetic field strength at the time of the flyby is not known. We list in Table 1 typical values of the interplanetary magnetic field (IMF) magnitude (|**B**|) at this distance in the solar wind if 0.1-0.3 nT [33, 24]. If we assume the interplanetary shock that passed New Horizons on DOY 190 also increased |**B**| to the top of this range, we calculate an Alfvén speed ($V_A$ = B/($\mu_o$ ρ)$^{1/2}$) of 41 km s$^{-1}$, an Alfvén Mach number ($M_A$ = $V_{SW}/V_A$) of 9.8, and magnetosonic Mach number of 9.5, reduced to $M_{MS}$=8 if we include the IPUIs. The ratio of proton thermal pressure to magnetic pressure ($\beta$ = nkT/[B$^2$ /2$\mu_o$]) is just 0.07, which would suggest the IMF dominates the dynamics. But if we include the substantial pressure of the IPUIs $\beta$ becomes greater than unity, emphasizing the importance of IPUIs for the dynamics of the outer heliosphere. We stress that all of these values are based on assumed field strengths and not measured values.

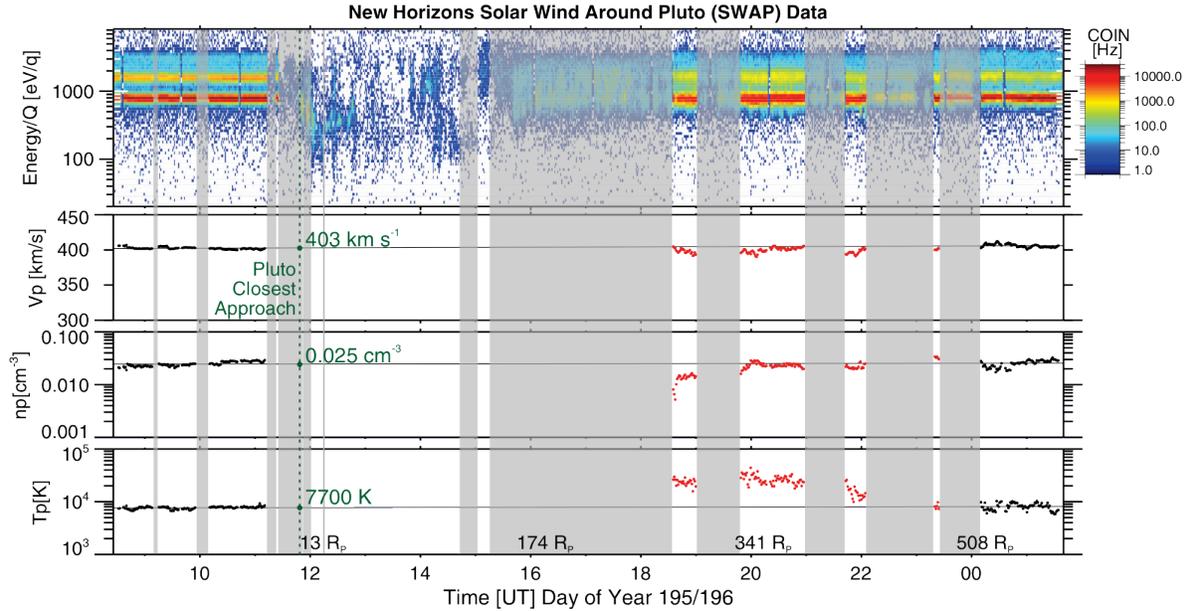

***Fig. 2. Overview of SWAP data.*** *Color spectrogram of coincidence count rate (COIN) as a function of E/q and derived proton flow speed, density, and temperature values for the interval surrounding the Pluto flyby. The vertical dashed green line shows the time of closest approach. Unshaded regions indicate times during which some portion of SWAP's very broad field-of-view was pointed within 5° of the Sun direction, and thus SWAP would be able to observe a radially outflowing solar wind; moments are derived for these times when solar wind-like distributions were observed. Black data points at the start and end of this interval allow linear least squares fits to these samples of essentially pristine solar wind (nearly horizontal lines), while disturbed, higher temperature solar wind (red points) are evident in the interaction region several hundred $R_P$ behind Pluto.*

## Upstream Interaction Confined Close to Pluto

Heavy ions picked up sunward from Pluto should mass-load and slow the solar wind ahead of it. However, there is no evidence of such solar wind slowing, and hence no evidence of the addition of Pluto pick-up ions (PPUIs) in the SWAP data as close in as ~20 $R_P$ inbound (point A in Fig. 1 at 11:25 UTC). The calculated speed value at that location is 405 km s$^{-1}$, but if we take an upper bound of 1% slowing at this point, or 399 km s$^{-1}$ compared to the interpolated value at Pluto of 403 km s$^{-1}$, we can use conservation of momentum to calculate an upper bound on the density of fully picked up heavy ion (CH$_4$ [2]) at this distance. We note that if the pickup ions are N$_2$ instead of CH$_4$, then densities would be smaller by the mass ratio of the ions (28/16) or 1.75. For a solar wind proton density of 0.025 cm$^{-3}$ we obtain

$$403 \text{ km s}^{-1} \times 0.025 \text{ cm}^{-3} = 399 \text{ km s}^{-1} \times 0.025 \text{ cm}^{-3} (1 + 16 N_{CH4}/N_P) \qquad \text{Eqn. 1}$$

providing an upper bound on the $N_{CH4}/N_P$ ratio of ~$6\times10^{-4}$ and upper bound on the fully picked up $CH_4$ density of $2\times10^{-5}$ cm$^{-3}$ at 20 $R_P$ along the New Horizons trajectory.

Between 11:25 and 11:56 UTC the New Horizons spacecraft was pointed in directions that did not allow SWAP to view back toward the Sun and into the solar wind. However, from 11:56-11:58 UTC (point B in Fig. 1) the spacecraft rotated through directions sufficiently close to the sunward direction that SWAP was able to take three contiguous energy-per-charge scans of what appears to be solar wind plasma. For each sample, we made Gaussian fits around the peak of the coincidence counts, including errors, and calculated speeds of 314, 343, and 315 km s$^{-1}$, for an average of 324 km s$^{-1}$. Therefore, we conclude that there was about a ~20% slowing of the solar wind at this location. Repeating the conservation of momentum calculation as above

$$403 \text{ km s}^{-1} \times 0.025 \text{ cm}^{-3} = 324 \text{ km s}^{-1} \times 0.025 \text{ cm}^{-3} (1 + 16 N_{CH4}/N_P) \qquad \text{Eqn. 2}$$

produces an approximate $N_{CH4}/N_P$ ratio of ~$2\times10^{-2}$ and fully picked up $CH_4$ density of ~$4\times10^{-4}$ cm$^{-3}$ at this location along Pluto's dawn flank. This very small amount of mass loading so close in to Pluto demonstrates that Pluto cannot have a strong comet-like interaction as was generally thought before the flyby.

*Extended Interaction Behind Pluto*

For roughly three hours after closest approach to Pluto, SWAP observed much lower levels of coincidence counts and no obvious beam-like distribution characteristic of the solar wind. From ~12:00-14:40 UTC and again from ~15:00-15:15 UTC (unshaded regions in Figure 2), SWAP was viewing close enough to the sunward direction that it would have seen such a beam. However, because (i) a relatively narrow, solar wind-like distribution could have been deflected in this region, and (ii) sufficiently slow flowing solar wind plasma would drop below the energy range sampled by SWAP, it is not possible to determine from these coincidence data if the solar wind was somehow excluded from all or part of the sampled region and/or if it was simply flowing in a way that SWAP was not able to observe. More detailed analysis of the SWAP data, including identification of the light (solar wind) versus heavy (Pluto) ions will be required in order to understand the plasma distributions observed behind Pluto.

Starting at ~18:30 UTC, New Horizons was turned several times to observe the solar wind direction and SWAP measured disturbed solar wind with somewhat variable speed and density and proton temperatures up to ~40,000 K (3.4 eV), much higher than the ~7700 K (0.7 eV) of the surrounding solar wind. As seen in Figure 2 (red points), the temperature shows elevated values that generally drop off with distance over 400 $R_P$ and return to the essentially unperturbed values by around the start of DOY 196. These SWAP observations suggest significant heating by the Pluto interaction. Finally, by the time that New Horizons next viewed the solar wind at the start of DOY 196, the plasma conditions had essentially returned again to those observed before the flyby, indicating the end of any significant Pluto interaction.

**Particle Measurements**

*The PEPSSI Instrument*

The PEPSSI instrument measures the time-of-flight (TOF) of energetic ions and electrons by detecting their passage between start and stop foils. For those particles with sufficiently high energy, the energy deposited in a given solid state detector (SSD) is measured [4]. PEPSSI has six angular sectors labeled S0-S5, and we concentrate on the measurements in S0, which observes particles close to the Sun-direction during the time of the Pluto encounter. We report on TOF-only ion measurements corresponding to energy per mass of ~0.5-50 keV/amu, for which PEPSSI observed the highest counting rates at Pluto. The PEPSSI TOF range nominally extends from 3-168 ns [35] for time intervals reported for ions traversing a 6.00-cm path internal to the instrument.

Ions are accelerated into PEPSSI by a potential of -2.63 kV by a negatively biased grid at the entrance apertures. This potential is with respect to the spacecraft which we assume to be close to that of the ambient medium. The reported energy/mass range given above is measured inside the instrument. Lacking knowledge of an ion's mass, there is ambiguity about what ion species are being measured. The likely composition of ions in the solar wind in the outer heliosphere include $H^+$, $He^{++}$, $He^+$ and $O^{n+}$ either from the ions originating in the solar corona [36] or from IPUIs originating in the interstellar medium [37].

*PEPSSI Observations of the Pluto Environment*

Pluto has a significant effect on the interstellar pickup and suprathermal ions that pass Pluto or its downstream disturbance, as illustrated in **Figure 3**. We have found no decisive evidence for plutogenic heavy ions in the PEPPSI energy range in the immediate vicinity of Pluto (within 500 $R_P$ of Pluto). There were also no detections near Pluto of >25keV electrons above a background consistent with expected galactic cosmic ray fluxes.

Figure 3 shows an overview of the PEPSSI measurements obtained during the Pluto flyby. New Horizons entered the region of Pluto's interaction as seen in the energetic particles sometime between 06:20 and 11:52 UTC on the flyby day. PEPSSI data suggest New Horizons left that region sometime between 15:20 and 18:33 UTC, equivalent to a downstream distance of 120-270$R_P$. These periods are marked by green background shading in Figure 3. The energetic particles return to a normal state after 18:33 UTC. The earliest and latest times delimit periods using two criteria, when PEPSSI was close to its nominal attitude (sector S0 of the instrument pointed ~35° from the Sun direction) and when measured fluxes and spectra were characteristic of those observed in the interplanetary medium.

Around 11:57 UTC (point B in Fig. 1) and 50° from the solar direction, PEPSSI detected intensities >10 times larger than those typically observed for that attitude. New Horizons was at that time about 10 $R_P$ to the side of Pluto. **Figure 4** shows a close-up of this time period. The envelope of the measured intensities is modulated by the change in PEPSSI's look direction, and there is a superimposed fine structure on the timescale of several seconds (corresponding to ∼ 0.1 $R_P$). This fine structure together with the higher than normal intensities for this look direction suggests that New Horizons entered a new region. One hypothesis is that PEPSSI detects particles accelerated from lower energies at a compressive shock near Pluto. Such a shock might be similar to the exotic shock observed at Titan in the solar wind [38]. Moreover, the TOF spectrum at that time is too similar to measurements in the interplanetary medium (compare Figure 3) to make this scenario likely. Another hypothesis is that the intensity enhancements result from the flow being deflected closer to the PEPSSI field of view and that the fine structure reflects rapid changes associated with the flow being turbulent about this mean deflection. This interpretation has the virtue of simultaneously explaining the similarity of the spectra and difference in the direction relative to the Sun, as well as the position of the enhancement relative to Pluto.

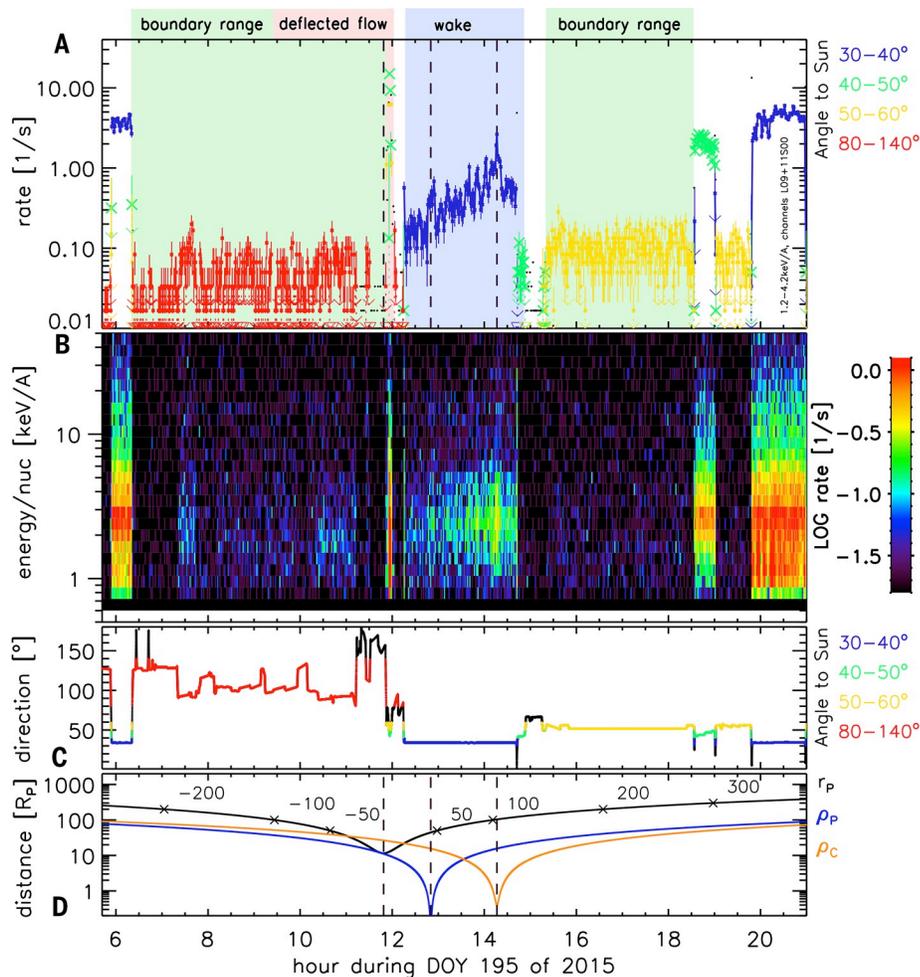

*Fig. 3 Overview of PEPSSI data taken near Pluto.* Panel A: Measurement of ions with Poisson error bars based on time of flight (TOF) data. The ions have speeds inside of the *instrument that correspond to 1-4keV/amu. The colors of the symbols identify the viewing angles relative to the Sun direction (panel D). Background colors refer to time periods discussed in the text. Dashed lines mark locations where New Horizons had its closest approach to Pluto and was directly anti-sunward of Pluto and Charon. Panel B: Energy spectrogram of TOF data assuming ions have not been accelerated by the potential, nor accounting for energy losses in the foils Panel C: Angle of PEPSSI's sector S0 to the Sun direction, with colors corresponding to those used in panel A. Panel D: Location of New Horizons relative to Pluto and Charon. Black: radial distance to Pluto, blue: distance along Pluto-Sun line, red: distance to Pluto-Sun line, orange: distance to Charon-Sun line.*

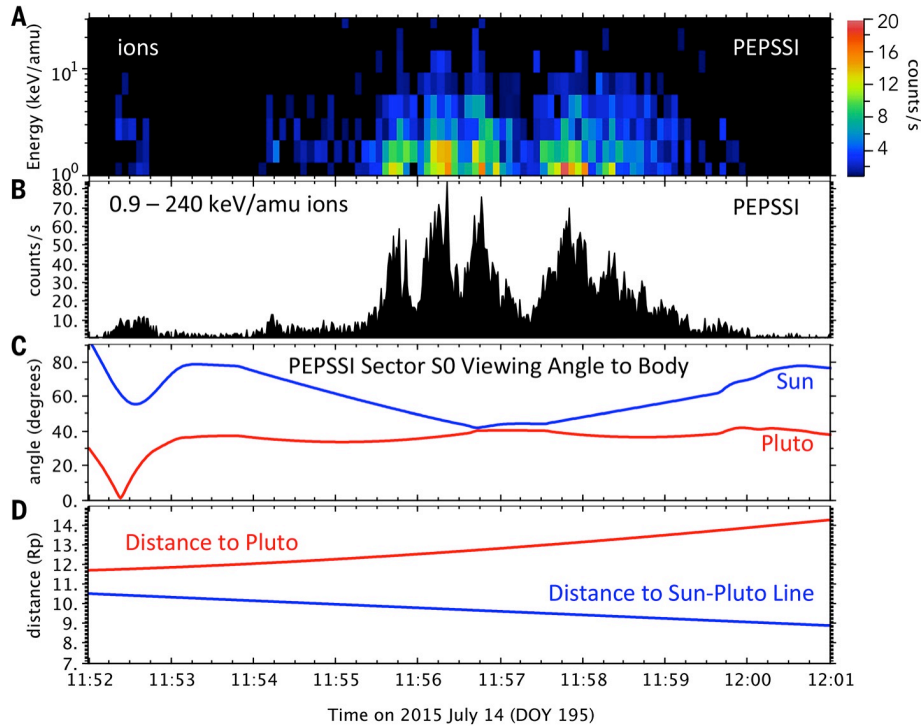

*Fig. 4: **Burst of energetic ions.** Zooming in to the interval shaded pink in Panel A of Fig. 3, we show PEPSSI TOF observations from 11:52 to 12:01 UTC on 14 July 2015. The top panel shows a color spectrogram of 6,625 single TOF events returned during this period from all PEPSSI sectors. Raw data are acquired once every second and have been binned into 5-sec average rates and logarithmically spaced bins in energy per mass. The second panel shows corresponding total counts per second integrated across all energies. The third panel shows the angle between the direction to the Sun and the normal to the S0 sector. The fourth panel shows the distance of New Horizons to Pluto and to the Sun-Pluto line during this period.*

From at least 12:15 to 14:42 UTC (to at least 100$R_P$ downstream, cf. Fig. 3), New Horizons was in a region very different from that observed in the interplanetary medium, which we assume to be a wake downstream of Pluto. Although PEPSSI was in its nominal attitude favorable for measuring pickup ions, the intensities at TOFs equivalent to ~1-4 keV/amu were ~10 times lower than in the surrounding interplanetary medium. We note that the intensity and not the spectral shape appear to be changing over time. The color scale along the trajectory curve in **Figure 5** shows a gradual increase in count rate during this period as the spacecraft departs from Pluto. It is distance from Pluto, not distance away from the Sun-Pluto line that organizes the intensification trend. The TOF observations show a break in the energy spectrum, at similar energies to instances in the interplanetary medium.

The ion intensity throughout the wake exhibits 20-minute quasi-periodic enhancements superposed on the overall trend. These might be a result of turbulent flow within the wake similar to the possible deflected flow interpretation of observations closer to Pluto, but the scale is much larger in the wake. Two of these enhancements are broader with sharp peaks in their center and coincide with New Horizons passing the geometric wakes of Pluto and Charon. Figures 3 and 5 show these enhancements at 12:50 UTC when New Horizons was 44 $R_P$ downstream of Pluto and at 14:17 UTC when 97 $R_P$ downstream of Charon. These enhancements are reminiscent of measurements at 12-25 lunar radii downstream of Earth's Moon immersed in the solar wind and magnetosheath [39]. At the Moon there is a density enhancement on the central axis of the wake where protons refilling the wake parallel and anti-parallel to the magnetic field merge together.

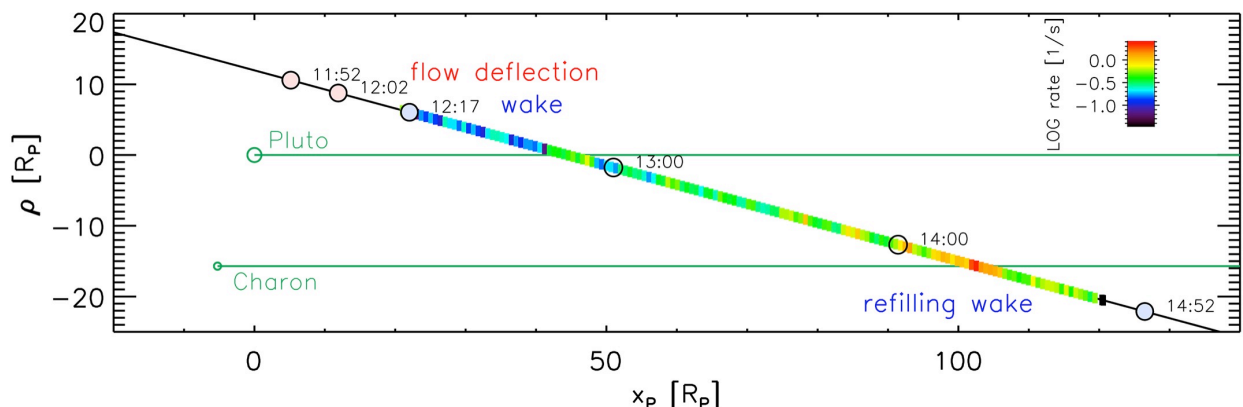

*Fig. 5: Energetic particles along the trajectory of New Horizons.* *The $x_P$-axis points away from the Sun. The vertical axis shows $\rho_P$, the distance from the $x_P$-axis in a plane containing Pluto and the trajectory. Green circles show the locations of Pluto and Charon. Open circles show times in UTC. The pink-filled circles mark the range of the ion enhancement; blue-filled circles delimit the wake. Color along the trajectory shows a PEPSSI count rate (same as in Fig. 3, Panel A).*

PEPSSI data reveal an interaction between Pluto and the solar wind with a scale of at least 11 $R_P$ on the flank of Pluto and extending at least 84 $R_P$ downstream in the solar wind when the spacecraft attitude changed to a non-favorable orientation for PEPSSI observations before completely exiting the Pluto interaction region.

**Dust Measurements at Pluto**

*The Venetia Burney Student Dust Counter*

The Venetia Burney Student Dust Counter (SDC) is an impact dust detector onboard the New Horizons spacecraft. SDC measures the mass of dust grains in the range of $10^{-12} < m < 10^{-9}$ g, covering an approximate size range of 0.5-10 $\mu$m in particle radius [5]. Since April 2006, SDC has been taking near-continuous measurements across the solar system [40, 41], and already provided estimates for the dust production rate and initial size distribution of dust in the Edgeworth-Kuiper belt [42, 43]. It is the first dedicated dust instrument to reach beyond 18 AU.

During the Pluto encounter the spacecraft executed a complicated sequence of attitude changes by firing its thrusters in order to optimize the observations. In order to avoid recording of excessive noise during the encounter, on January 1, 2015 we set our charge threshold to $Q \sim 10^7$ e, corresponding to a smallest detectable particle radius of 1.4 $\mu$m. This charge threshold is above the level of the majority of SDC recorded thruster events throughout the mission. The thresholds were reset to their pre-encounter values on DOY 211, 2015.

### *Dust Measurements During the Encounter*

A period of ± 5 days centered on closest approach, corresponding to approximately ± 5000 $R_P$ is used to calculate the Pluto system dust density distribution. SDC recorded a total of 102 events in this time period. Throughout the mission, due to the expected low dust fluxes, coincident events between multiple channels, or events coincident with thruster firings were identified as noise events for both the exposed and the reference detectors. **Figure 6** shows all the recorded events through the encounter.

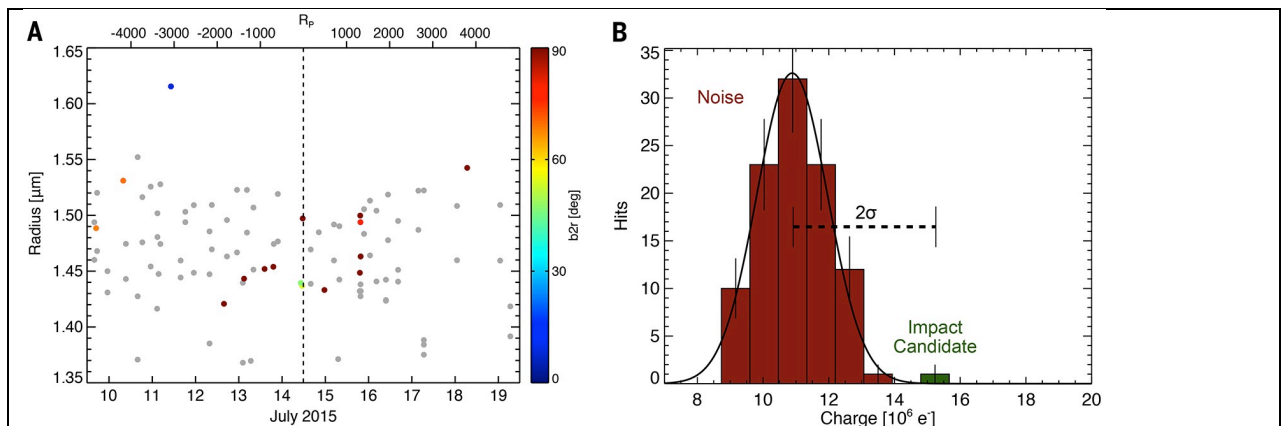

*Fig. 6 Events around Pluto. (A) The events recorded during ± 5 days of the closest approach to Pluto. The gray colored dots mark noise events that were identified to be coincident dust hits, or a single event coincident with thruster firings. The color-code represents the bore-sight-to-ram angle, measured between the SDC surface normal and the velocity vector of the*

*spacecraft. SDC's sensitivity rapidly drops to zero for impact angles > 45º, hence events marked in red/green colors are also noise events. A single detection at a distance of about 3000 Rp (July 11, 2015) before closest encounter remains the only candidate for detecting a Pluto-system dust particle. (B) The amplitude distribution of all the noise events during this period is well fitted with a Gaussian curve with an average of $1.1 \times 10^7$ e and a $1\sigma = 1.5 \times 10^6$ e, indicating that our candidate impact event generated a charge with an amplitude $2\sigma$ above the average.*

After identifying the coincident events as noise, 16 events remained as candidate dust hits. None of these events occurred on the two reference detectors. New Horizons passed through the Pluto system with a speed of $v_{s/c} \sim 14$ km/s relative to Pluto. Dust grains in the Pluto system are expected to have speeds $<<v_{s/c}$, hence $v_{s/c}$ becomes to a good approximation the impact speed. Due to the sequence of observations executed by New Horizons, the orientation of the spacecraft changed almost continuously during the encounter, pointing SDC only intermittently in the ram direction. Because of the impact angle dependence of SDC's sensitivity, the detection probability of dust particles with impact angles > 45° approaches 0, and these events are also identified as noise. Throughout the entire period of the close encounter SDC recorded only a single large amplitude event that could be due to a dust impact. To assess to probability that this event was not noise, we examined the amplitude distribution of the recorded noise events, and estimated that the detection is outside the $2\sigma$ error of the average amplitude of the noise events, hence it has a probability of ~95% to be dust particle. We use this single detection as an upper limit to estimate the dust density near Pluto (see SOM), indicating a most likely density of 1.2 km$^{-3}$ and a 90% confidence level for the density to be in the range of $0.6 < n < 4.6$ km$^{-3}$.

**Figure 7** compares this density estimate with dust measurements in the outer solar system, indicating that the dust density of particles with radii > 1.4 μm remained within 1 σ error bar of our last data point representing the SDC measurements since January 1, 2015.

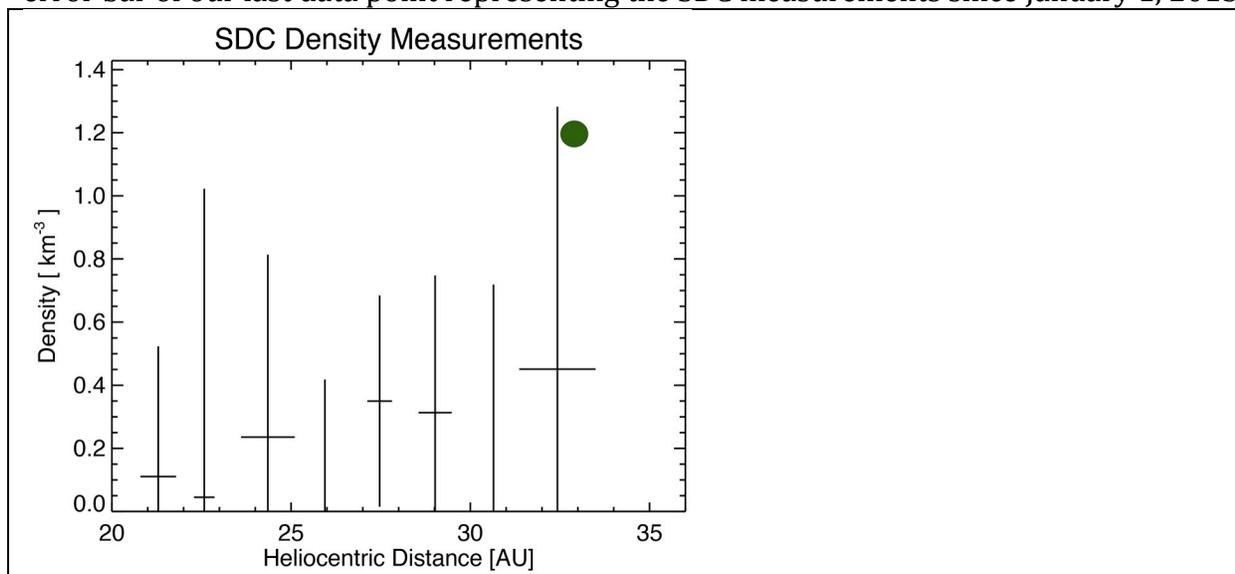

*Fig. 7. Dust in the outer solar system. The dust density or particles with radii > 1.4 μm measured by SDC in the outer solar system. The last data point with error bars shows the data collected since January 1, 2015. The green dot indicates the most likely dust density of*

*1.2 km$^{-3}$, based on a single candidate dust event.*

The plasma wave instruments on Voyager 1 and 2 also showed a roughly flat dust density of ~ 0.2 km$^{-3}$, though in that case the size of the detected particles remained poorly determined [44]. This match can now be used to estimate the Voyager dust detection threshold to be similar to SDC's during the Pluto flyby.

The dust density distribution perhaps indicates a slight increase with distance. This could be the result of the inner edge of the Kuiper belt dust disk extending inward and engulfing the outer solar system. Alternatively, the dust density increase could be local to the Pluto system. As SDC will map the dust density distribution for years to come, we will learn how the trend continues deep into the Kuiper Belt.

**Discussion**

***The Obstacle Pluto Presents to the Solar Wind***

Pluto presents a unique obstacle to the solar wind and any theory that seeks to explain it has to account simultaneously for a challenging range of observations provided by New Horizons. In particular, the SWAP observations of limited (<1%) slowing down of the solar wind at ~20 R$_P$ upstream of Pluto (point A in Fig. 1) suggest very few atmospheric molecules are escaping, becoming ionized and mass-loading the solar wind. When the spacecraft was 8.8 R$_P$ tailward from Pluto and at a transverse distance from it of 9.6 R$_P$ (point B in Fig. 1) the solar wind had slowed by ~20%. At this time PEPSSI detected an enhancement of keV ions. While there is no current consensus on the nature of the interaction boundary, we can still estimate its size. We can make a zeroth-order calculation of the size of the Pluto obstacle to the solar wind by assuming the same transverse distance applies at Pluto's terminator and multiplying by 2/3 (based on experience at terrestrial planets) to get a very approximate distance of ~6 R$_P$ for the 20% slowing location directly upstream of Pluto. This distance is about twice as large (scaled to the planet) as the maximum observed bow shock distance in the terminator plane for Mars and Venus [45]. The interaction at Pluto is therefore consistent with a Mars-like interaction, given the current uncertainties in its bow shock location.

The solar wind interaction with Pluto's atmosphere is expected to depend on (a) the solar wind flux at Pluto, which varies by a factor of ten on timescales of a few days and by $1/a^2$ with Pluto's heliocentric distance $a$; (b) the rate of escape of Pluto's neutral atmosphere; (c) the ionization rate of Pluto's atmosphere which, for both photoionization and charge-exchange, also varies as $1/a^2$. The enhanced solar wind pressure (due to recent passage of a interplanetary compression region) at the time of the Pluto encounter suggests that the interaction region was in a compressed state at the time of the New Horizons flyby.

If the "obstacle" is mass-loading the solar wind via ionization of an escaping atmosphere then we would expect the size of the obstacle to be inversely proportional to the upstream solar wind momentum flux. Comparing the observed solar wind flux with typical Voyager 2 values (Table 1), the factor of ~4 enhancement at the time of the New Horizons flyby suggests that more typical size range for the obstacle (for the same atmospheric escape rate) would be ~25 $R_P$.

On the other hand, if the solar wind does not suffer significant mass loading due to ionization of an escaping neutral atmosphere well upstream of the object, then the solar wind will be slowed by ion pick-up (and subsequent mass-loading) in the outer atmosphere and diverted by electrical currents induced in Pluto's ionosphere. The size of such an interaction, similar in nature to Mars and Venus, is set by the altitude of the peak ionospheric electron density and how sharply the atmospheric density drops with altitude. This Mars-type interaction of the solar wind with an exosphere/ionosphere would be less compressible and fluctuate less in size with solar wind flux.

While the small size of the interaction region relative to the Pluto is reminiscent of Mars and Venus, we note that recent observations of the solar wind interaction with the relatively weakly out-gassing Comet 67P Churyumov-Gerasimenko by instruments on ESA's Rosetta spacecraft [46, 47, 48] show deflection of the solar wind with relatively modest decrease in speed. We anticipate interesting scientific discussions of the relative roles of atmospheric escape rate, solar wind flux, and IMF strength at Mars, Comet 67P and Pluto as the data from the MAVEN, Rosetta and New Horizons spacecraft are further analyzed.

*Atmospheric Escape*

Pluto's atmosphere was first detected in 1988 during stellar occultation [49] and later determined to be primarily composed of $N_2$ with minor abundance of $CH_4$ and CO, with surface pressures of ~17 microbar [50, 51]. Pluto's low gravity implies that a significant flux of atmospheric neutrals can escape. Estimates of escape rates range from as low as $1.5 \times 10^{25}$ s$^{-1}$ to as high as $2 \times 10^{28}$ s$^{-1}$. The most recent (pre-New Horizons) atmospheric model of [23] indicates a denser and more expanded atmosphere with an escape rate of ~$3.5 \times 10^{27}$ $N_2$ s$^{-1}$ and an exobase at 8 $R_P$~9600 km. These are the conditions that were anticipated on arrival at Pluto.

The New Horizons trajectory was designed to provide solar and Earth occultations of Pluto's atmosphere by the UV spectrometer (Alice) and Radio Experiment (REX) instruments respectively [52, 53]. These occultation measurements revealed Pluto's atmosphere to be colder and less extended than expected [2]. Matching the Zhu et al. [2014] [8] model to the New Horizons data suggests cooler atmosphere with an exobase height of 2.5 $R_P$ and an escape rate of only $6 \times 10^{25}$ molecules s$^{-1}$ comprising mostly methane (rather than nitrogen). This limited atmospheric escape drastically reduces the neutral material upstream of Pluto available for ionization and mass-loading the solar wind.

*Ionosphere*

New Horizons did not make a direct measurement of Pluto's ionosphere. Adapting a pre-encounter model of the atmosphere to the density, temperature and composition measurements from New Horizons [2], we find a peak electron density of ≤1300 cm$^{-3}$ at a distance of 1900 km =1.6 R$_P$. The density drops above this peak with a scale height of ~330 km. The main ions are H$_2$CN$^+$ (mass=28 amu) and C$_2$H$_5^+$ (mass=29). Preliminary estimates of the electrical conductivity of such an ionosphere suggest it is sufficient to sustain currents that would divert the solar wind.

*Bow Shock*

In modeling interactions – cometary or Mars-like - it is often assumed that the planet's atmosphere/ionosphere and the solar wind can be considered as fluids. For many solar system bodies fluid descriptions of a plasma-obstacle interaction are often good starting points. Global-scale magnetohydrodynamic (MHD) models have been successful in capturing the basic structure of many plasma interactions. The fast, cold flow of the solar wind in the heliosphere is highly super-sonic (see Table 1). We illustrate the shape of a Mach 40 shock in Figure 1 to show how bent such a shock could be behind Pluto. We also show a low Mach number shock to illustrate how including a high density of IPUIs and/or substantial upstream ionization of plutogenic PPUIs would move the shock farther upstream as well as reduce the shock angle.

With the IMF being very weak at Pluto's orbital distance, the length scales on which the plasma reacts are large compared with the size of the interaction region. For instance, at 33 AU the gyroradius of solar wind protons is ~23 R$_P$ and the pick-up ion gyroradius of CH$_4^+$ ions is ~ 200-800 R$_P$ [7, 9].

There is no direct evidence of a sharp bow shock in Figs. 2 and 3. This may be because the spacecraft attitude was unfavorable during the passage of the shock (see the grey shading in Fig. 2 and the large angles in Fig. 3) so that it could not be observed well. However, it is important to consider the expected shock thickness. Two thicknesses that have been used in the past are the proton "turn-around" distance and the ion inertial length. For relatively strong magnetic fields, the turn-around distance is approximated by $V_{sw}/\Omega_{ci}$, which is proportional to $V_{sw}/B$ [54]. High Mach number shocks observed by Voyager at Uranus [55] and Neptune [56], as well as by Cassini at Titan [38] indicate this is a reasonable approximation with observed shock widths being ~30-70% of this quantity calculated from upstream conditions. For the observed upstream solar wind speed and |B| of 0.3 nT, we obtain ~50 R$_P$ for a solar wind proton turn-around distance. Perhaps a better scaling for the weak IMF conditions and small obstacle size at Pluto, for which the interaction may be mediated by whistler waves rather than shock-forming MHD modes, is the ion inertial length [54]. Again using the measured upstream density of ~0.025 cm$^{-3}$, we get an ion inertial length of ~1.2 R$_P$. Since the derived 20% slowing interaction distance of ~6 R$_P$ is intermediate between the scaled turn-around and ion inertial lengths, we conclude that an ionospheric obstacle could produce the derived dayside size scale of the solar wind slowing ahead of Pluto.

**Conclusions**

Pluto continues to deliver surprises. The New Horizons instruments that measure plasma and particles revealed an interaction region unlike any other body in the solar system and considerably smaller than predicted. This reduced interaction region is possibly due to a combination of a much smaller atmospheric escape rate than expected, indicated by the New Horizons atmospheric measurements [2] as well as the flyby occurring during a time of particularly high solar wind flux.

1. Observations indicated enhanced upstream solar wind flux detected by the SWAP instrument. The lack of any slowing until New Horizons was within 20 $R_P$ indicates that almost no heavy ions were ionized within the several thousand $R_P$ upstream of Pluto. The SWAP data revealed a surprisingly small interaction region, confined on its upwind side to within ~6 $R_P$ of Pluto. The interaction persists to over 400 $R_P$ behind Pluto.

2. PEPSSI has not found evidence of plutogenic pickup ions nor energetic electrons in its energy range within 500 $R_P$ of Pluto, but the interplanetary energetic particle intensities are significantly perturbed by the interaction. Changes in PEPSSI measurements near Pluto's terminator suggest that <10-keV ions are accelerated and/or deflected away from the direction radially from the Sun. PEPSSI observed decreased suprathermal particles in the wake of the interaction region. The particle intensities near Pluto decreased by a factor ~10 below the heliospheric value and increased with distance downstream.

3. During the encounter, SDC could detect grains with an effective radius greater than approximately 1.4 $\mu$m. Eliminating spurious events, such as thruster firings (leading to events which are not dust impacts) SDC detected 1 candidate impact in ±5 days around closest approach. In this time period, the effective volume SDC carved out is 0.83 km$^{-3}$, leading to a most likely dust density estimate of 1.2 km$^{-3}$, with a 90% confidence level range of 0.6 < n < 4.6 km$^{-3}$.

**ACKNOWLEDGEMENTS**

We thanks the many contributors to the development of the SDC, SWAP, and PEPSSI instruments and acknowledge many useful discussions with colleagues. The New Horizons mission is supported by NASA's New Frontiers Program. As contractually agreed to with NASA, fully calibrated New Horizons Pluto system data will be released via the NASA Planetary Data System at https://pds.nasa.gov/ in a series of stages in 2016 and 2017 as the data set is fully downlinked and calibrated.